\documentclass[a4paper,11pt]{article}
\pdfoutput=1 % if your are submitting a pdflatex (i.e. if you have
\usepackage{jinstpub} % for details on the use of the package, please
\usepackage{amsmath,amsfonts,amssymb}
\usepackage[latin1]{inputenc}
\usepackage{hyperref}
\usepackage{url}
\title{Use of the Peak-Detector mode for gain calibration of SiPM sensors with ASIC CITIROC read-out }

\collaboration{\includegraphics[height=13mm]{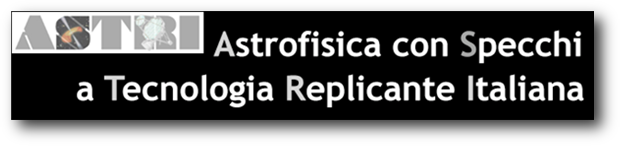}\\}
\author[a,1]{D. Impiombato\note{Corresponding author.},}
\author[a,2]{A. Segreto\note{Corresponding author.},}
\author[a]{O. Catalano,}
\author[a]{S. Giarrusso,}
\author[a,3]{and T. Mineo\note{Corresponding author.}}
\author[]{on behalf of the ASTRI Project}

\affiliation[a]{INAF, Istituto di Astrofisica Spaziale e Fisica Cosmica di Palermo, via U. La Malfa 153, I-90146 Palermo, Italy \\ }

\emailAdd{Domenico.Impiombato@inaf.it, Alberto.Segreto@inaf.it, Teresa.Mineo@inaf.it}

\abstract{The Cherenkov Imaging Telescope Integrated Read Out Chip (CITIROC) is a 
32-channel fully analogue front-end ASIC dedicated to the 
read-out of silicon photo-multiplier (SiPM) sensors that can be 
used in a variety of 
experiments with different applications: nuclear physics, medical 
imaging, astrophysics, etc. It has been adopted as front-end for the 
focal plane detectors of the ASTRI-Horn Cherenkov telescope \cite{pareschi16,scuderi18} and, in this 
context, it was modified implementing the peak detector reading mode to 
satisfy the instrument requirements. 

For each channel, two parallel AC coupled voltage preamplifiers, one for the 
high gain and one
for the low gain, ensure the read-out of the charge from 160 fC to 
320 pC (i.e. from 1 to 2000 photo-electrons 
with SiPM gain = 10$^{6}$, with a photo-electron to noise ratio of 10). 
The signal in each of  the two preamplifier chains is shaped and the  
maximum value is captured by activating the peak detector for an adjustable time interval. 

In this work, we illustrate the peak detector operation mode and, in 
particular, how this can be used to calibrate the SiPM gain without the need 
of external light sources. To demonstrate the validity 
of the method, we also present and discuss some laboratory measurements.

}


\begin{document}

\maketitle
\flushbottom
\section{Introduction}
\label{Introduction}
The calibration of the signals of any Cherenkov camera
is of considerable importance 
for the proper reconstruction of the recorded images.
A suitable calibration procedure involves measurements of the pixel electronics 
baseline, the linearity, the
uniformity of the response to incident light, and the value of 
the equivalent photo-electrons (pe), typically expressed in terms 
of analog-to-digital (ADC) counts.
Usually silicon photo-multipliers (SiPMs) have a good amplitude
 resolution showing a peaked 
distribution for multiple photo-electron entries, thanks to 
a very low excess noise factor resulting from the Geiger mode 
operation and a high uniformity in the response of the individual 
micro-cells of each SiPM pixel.  
One of the techniques that allows the gain calibration of a Cherenkov camera 
without the need of an external light source is based on dark acquisitions runs. 
In dark spectra one can observe a percentage of events with 
amplitudes larger than one photo-electron because of the so-called 
optical cross-talk effect. Secondary photons \cite{mirzoyan09} 
emitted in the primary avalanche can reach neighboring micro-cells, 
either directly or after a reflection, and trigger avalanches 
almost simultaneously to the primary one. This gives rise to high-amplitude pulses equivalent 
to the one produced by several incident photons. 
The dark spectrum shows therefore several peaks 
(with a Gaussian shape, to a good approximation) where
the distance between two neighboring peaks corresponds to the pixel gain.  
The drawback of this calibration technique is that, due 
to the low cross talk probability (<10\%), 
long acquisition time are necessary to accumulate spectra with
sufficient statistics in the peaks for an accurate determination of pixel gains.
This can be solved 
by means of a proper use
the peak detector mode implemented in CITIROC chip.

\newpage
\section{The CITIROC chip}
\label{The CITIROC chip} 
The CITIROC \cite{fleury14,impiombato15}, designed by WEEROC\footnote{http://www.weeroc.com}, 
is a 32-channel fully analogue front-end ASIC dedicated to the 
SiPM sensors read-out. 
It is designed with a 0.35$\mu$m SiGe 
technology from AustriaMicroSystems. It is an evolution of the EASIROC 
(the Extended Analogue Silicon photo-multiplier Integrated Read Out 
Chip)\cite{callier11,impiombato12,impiombato13,marano13,impiombato14} 
which was available since 2010 and used in a variety of 
nuclear physics and astrophysics and medical imaging applications. 
Fine-tuning of each pixel gain is obtained adjusting the voltage applied to the
SiPM through an 8-bit Digital-to-Analog Converter (DAC) ranging from 0 to 4.5 V.
For each channel, two parallel AC coupled 
voltage preamplifiers the High Gain (HG) and Low 
gain (LG) electronics chains ensure the read out of the charge from 160 fC to 
320 pC (i.e. from 1 to 2000 photoelectrons with SiPM Gain = 10$^{6}$, with a 
photo-electron to noise ratio of 10). 
 Two operation modes are implemented in the 
ASIC: the peak detector and the switched capacitor array (SCA).
The peak detector mode provides the maximum pulse height reached by the signal 
in a predefine time window (plus an 
electronics offset from the circuit), whereas 
the SCA mode returns the level of the shaped pulse height at the given 
sampling time. 
The trigger path that can be switched between either the LG or HG 
preamplifiers is composed of a 15 ns peaking time fast shaper followed 
by two discriminators. One discriminator provides trigger and hit 
information, with a programmable mask to block unwanted channels, the 
other provides accurate time information. An internal 10-bit threshold 
common to all 32 channels can be set for each of the two discriminators. 
Fine-trimming of each channel's discriminator is also possible with a 
4-bit DAC. The 32 analogue channel paths are multiplexed into one output 
each for HG and LG.

\section{The Peak detector mode}
\label{The Peak detector mode}

\begin{figure*}[ht!!]
\centering
\includegraphics[angle=0, width=12cm]{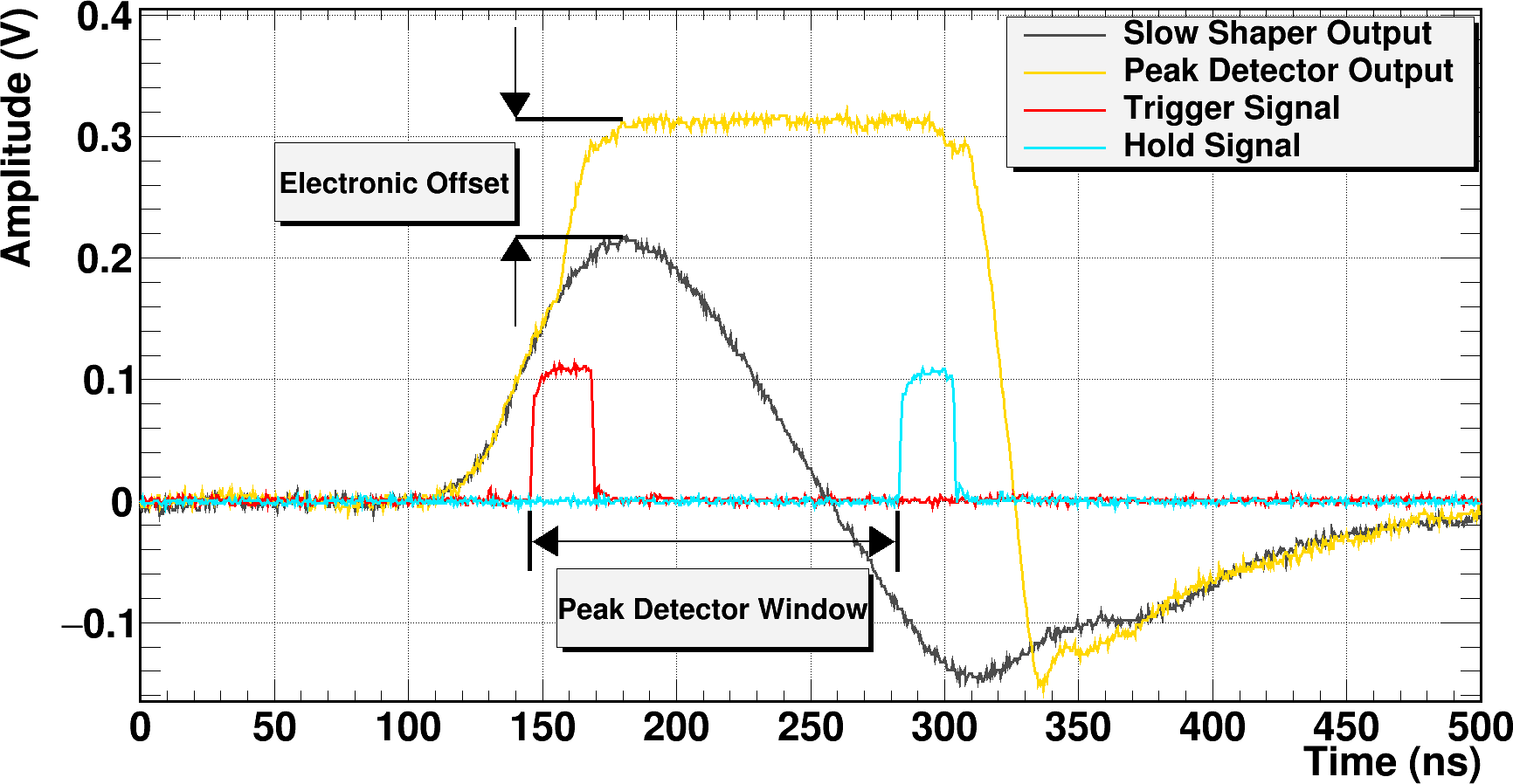}
\vspace{0.2cm}
\caption{Analog output  of CITIROC signals in peak detector working mode.
The black curve shows a SiPM pulse signal after the slow shaper as observed  
 in Track\&Hold operation mode. The yellow curve shows the output signal observed when the peak detector 
circuit is first activated by a trigger signal (the red curve) and then disabled by the 
falling edge of the hold signal (the blue curve).}

\label{fig1}
\end{figure*}

The peak detector working mode implemented in CITIROC is illustrated in 
Fig.~\ref{fig1}. The signal is generated by the SiPM is shaped with a time constant 
that can be varied from 12.5 to 87.5 ns, depending 
on the user requirements (black curve in Fig.~\ref{fig1}). The peak detector is activated, 
within a few ns, by a "trigger signal" (red curve in Fig.~\ref{fig1}) and is stopped after a 
predefined selectable delay, by the rising edge of the "hold signal" (blue curve in Fig.~\ref{fig1}).
During the "hold" signal the ADC conversion of the 
peak detector output is performed. 
The falling edge of the "hold" signal switches the chain back to track phase for the next 
acquisition. 
The delay between the trigger and the hold signals can be 
varied up to a few microsecond depending on the application of the ASIC. 

\newpage
\section{Peak detector output}
\label{Peak detector output}
\begin{figure*}[ht!!]
\centering
\includegraphics[angle=0, width=12cm]{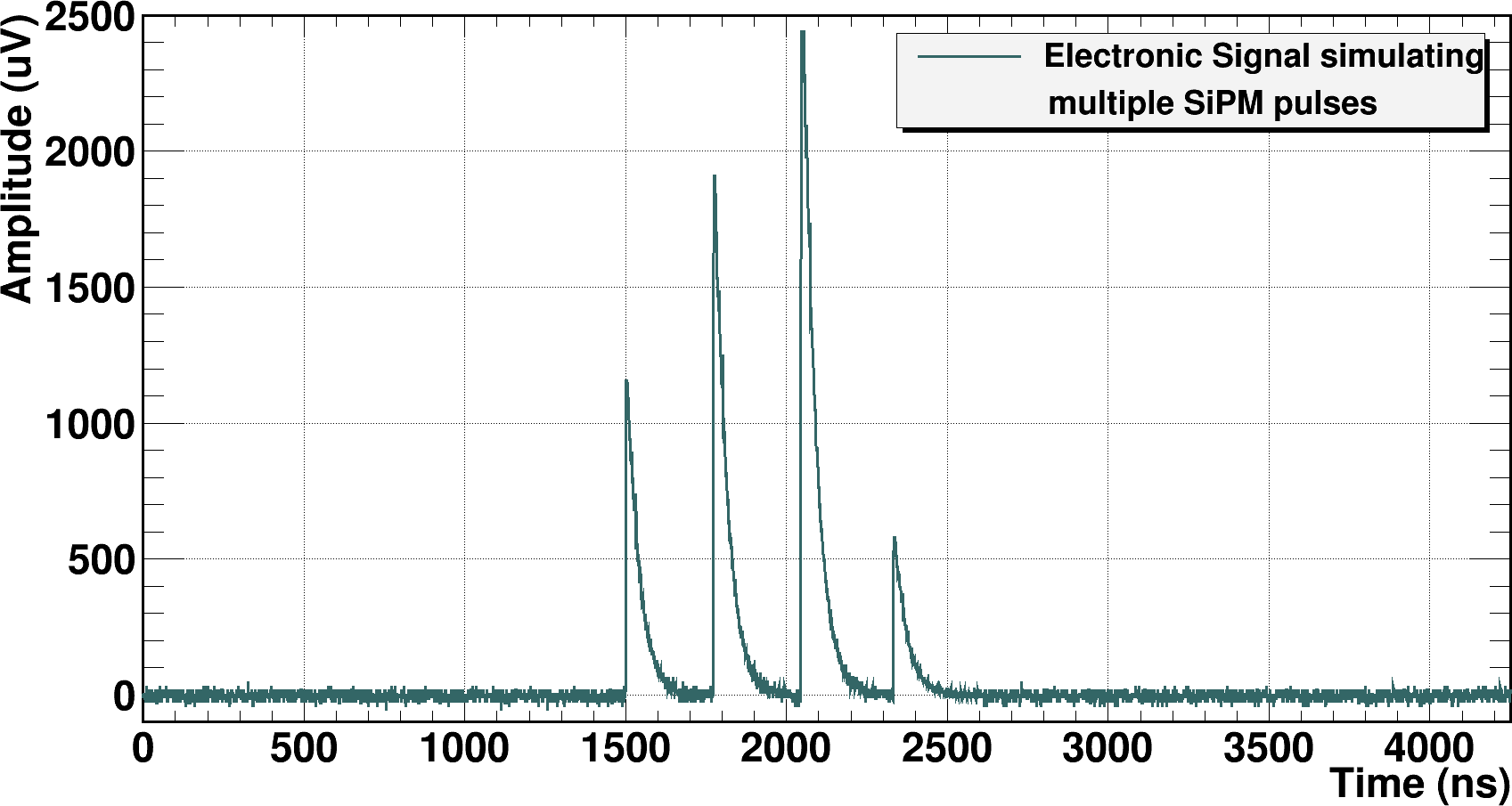}
\vspace{0.2cm}
\caption{Electronic signal injected into an ASIC channel to study the peak detector 
output in presence of multiple SiPM pulses detected within short time distance.}
\label{fig2}
\end{figure*}

\begin{figure*}[ht!!]
\centering
\includegraphics[angle=0, width=12cm]{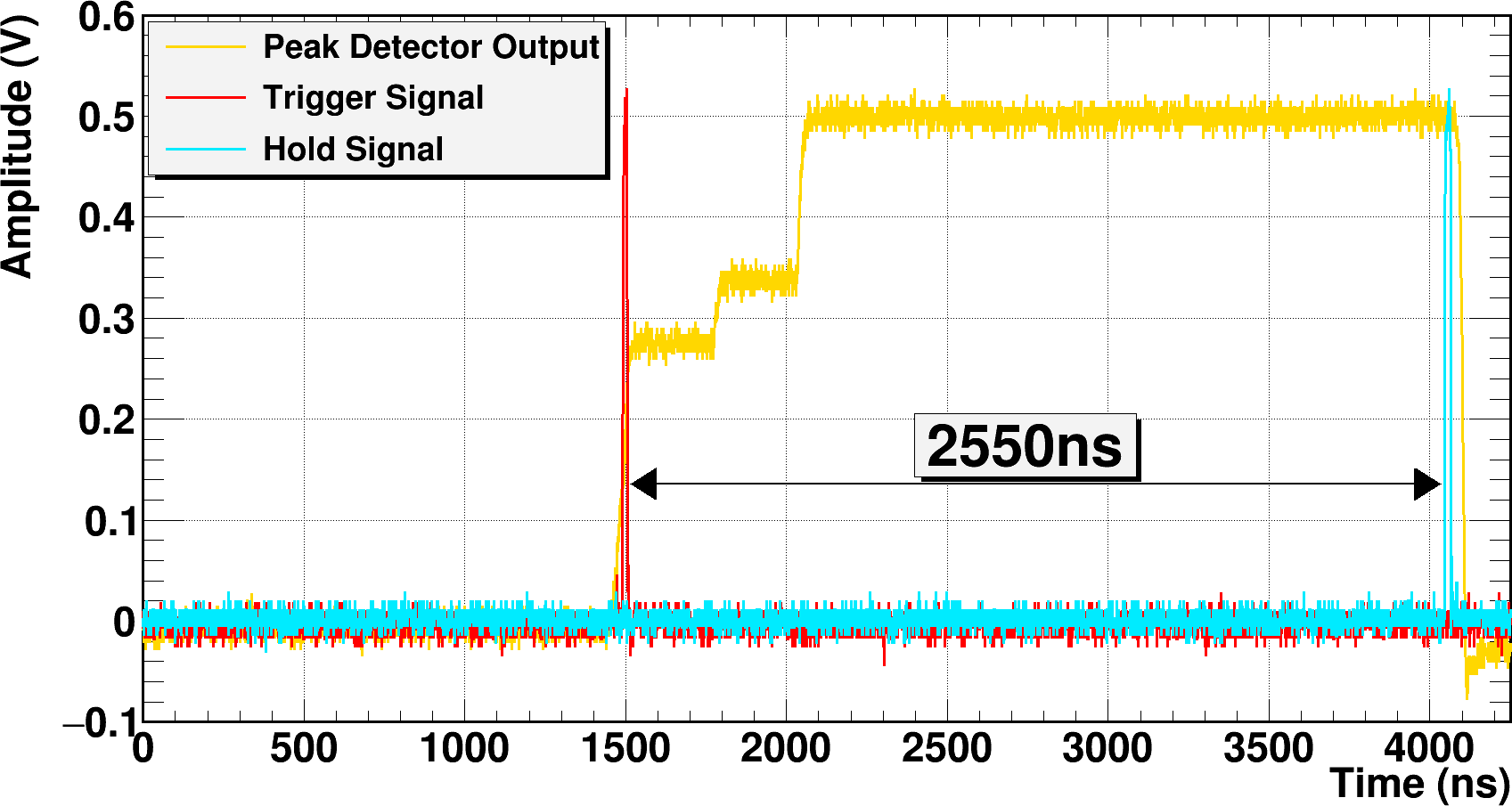}
\vspace{0.2cm}
\caption{Peak detector analog output when an electronic 
signal with multiple pulses (having different amplitudes) is injected.
The yellow curve shows the analog signal of the peak detector.
The red and blue curves are the trigger and hold signals, injected from the pulse generator.
As evident, after the injection of the highest amplitude pulse, the peak detector output 
stay at a perfectly constant level until the Hold signal is generated.}
%%The amplitude scale of the trigger and hold signals are normalized to the amplitude of the 
%%Slow Shaper HG signal.
\label{fig3}
\end{figure*}

\begin{figure*}[ht!!]
\centering
\includegraphics[angle=0, width=12cm]{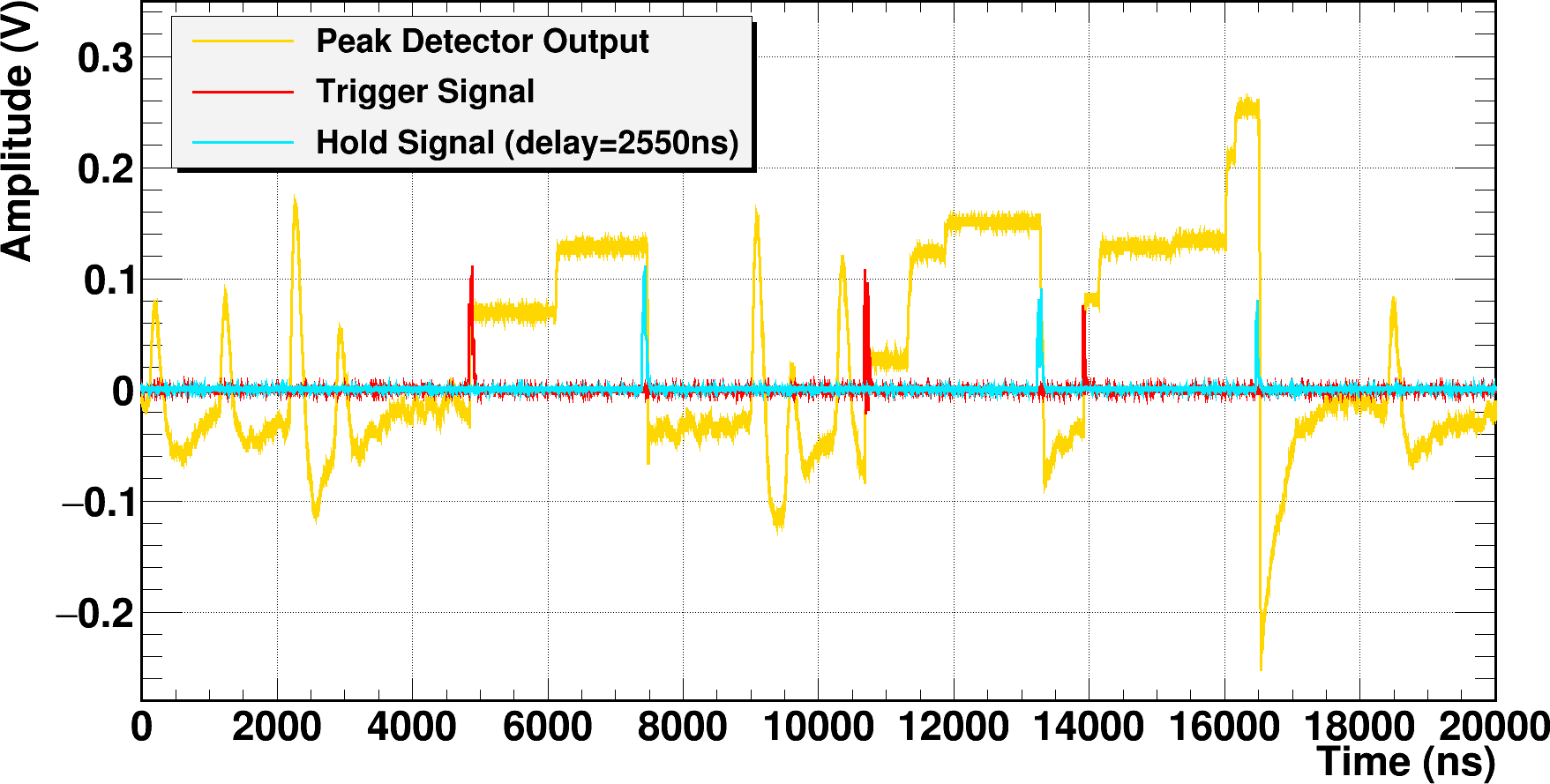}
\vspace{0.cm}
\caption{Peak detector analog output with real SiPM pulses in dark condition.
The yellow curve represents the  peak detector output, while
the red and blue curves are the trigger and hold signals, respectively.
The delay between trigger and hold signal has been set to 2550 ns.
In the peak detector signal, it is evident the detection of several pulses with amplitude higher than 1 pe 
generated by the cross-talk effect. } 
\label{fig4}
\end{figure*}

To study the characteristics of the peak detector output when multiple pulses 
at close time distance are generated by the SiPM,
we injected into one channel 
of CITIROC an electronic signal simulating four pulses, 
with different amplitudes, by using a 
pulse-function generator. The shape of the four pulses was tailored to the shape 
of a real SiPM output pulse i.e. with a very fast rise time 
(a few hundreds of ps) 
followed by an exponential decay with a time constant of about 150 ns (see 
Fig.~\ref{fig2}). The trigger and hold signals, provided by the pulse 
generator, were used to activate/deactivate the peak sensing circuit of the 
ASIC. The delay of the hold signal was set to a value of about 2550 ns
so that all the four simulated pulses were inside the 
peak detector activation time window. As shown in  Fig.~\ref{fig3}, 
the peak detector output depends only on the pulse with the highest amplitude.
In Fig.~\ref{fig4} we show the peak detector analog output as observed 
with real SiPM pulses generated in dark conditions.

\clearpage
\newpage

\section{Dark count spectra}
\label{Dark count spectra}

\begin{figure*}[h!!]
\centering
\includegraphics[angle=0, width=12cm]{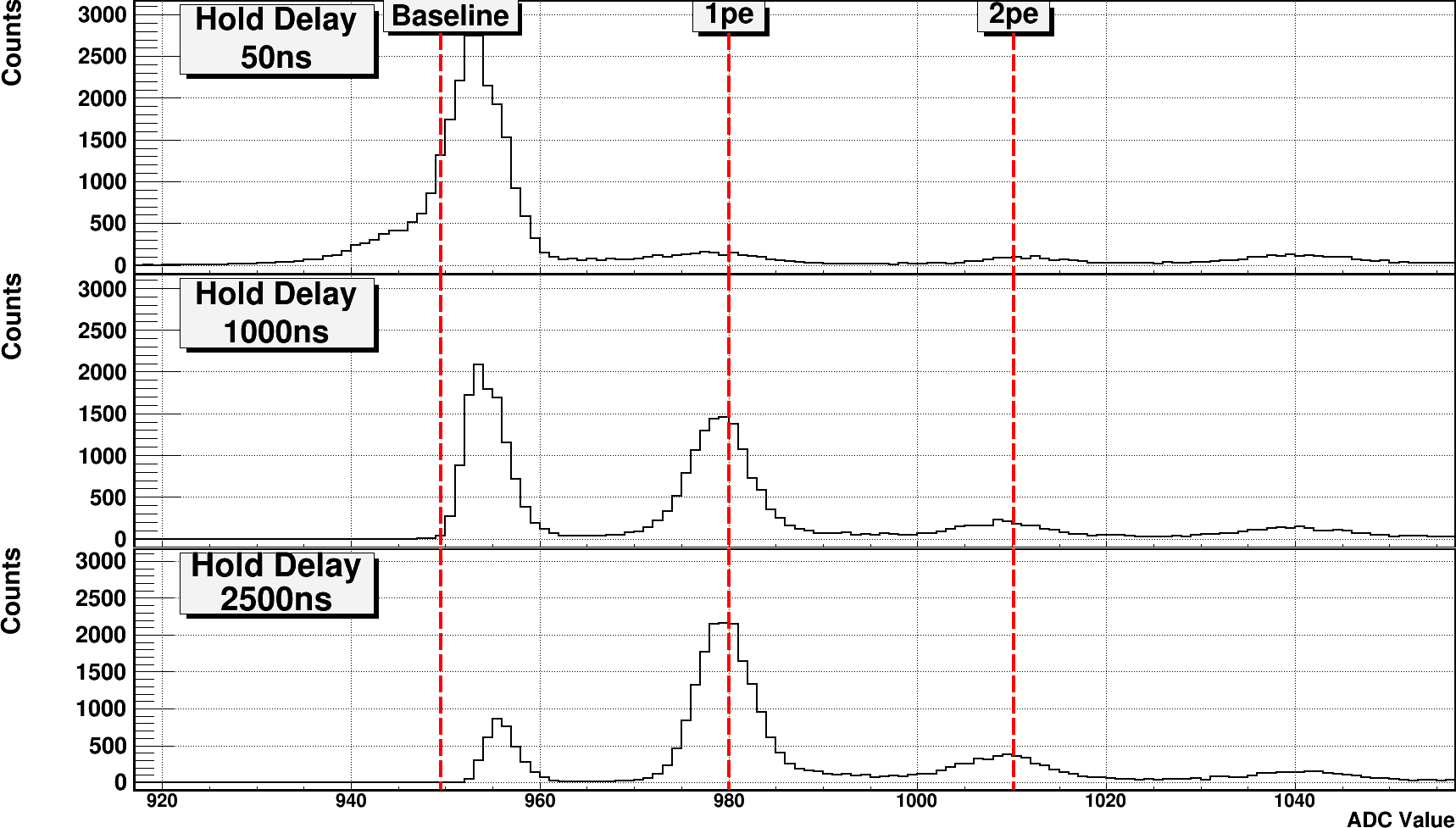}
\vspace{0.cm}
\caption{Example of the pulse height distribution of SiPM 
signals observed with an hold delay of 50 ns,
,1000 ns, 2500 ns, in dark condition. 
The red dashed line indicates the peak positions of the spectra relative 
to 1pe and 2pe.
The 0pe counts are not expected to be gaussian centered on the baseline due the fact that when 
no SiPM pulses are detected, the peak detector output will depend on the highest noise fluctuaction
in the peak detetector activation window. Therefore its statistical distribution will be biased toward values higher
than the baseline.}

\label{fig5}
\end{figure*}

\begin{figure*}[ht!!]
\centering
\includegraphics[angle=0, width=15cm]{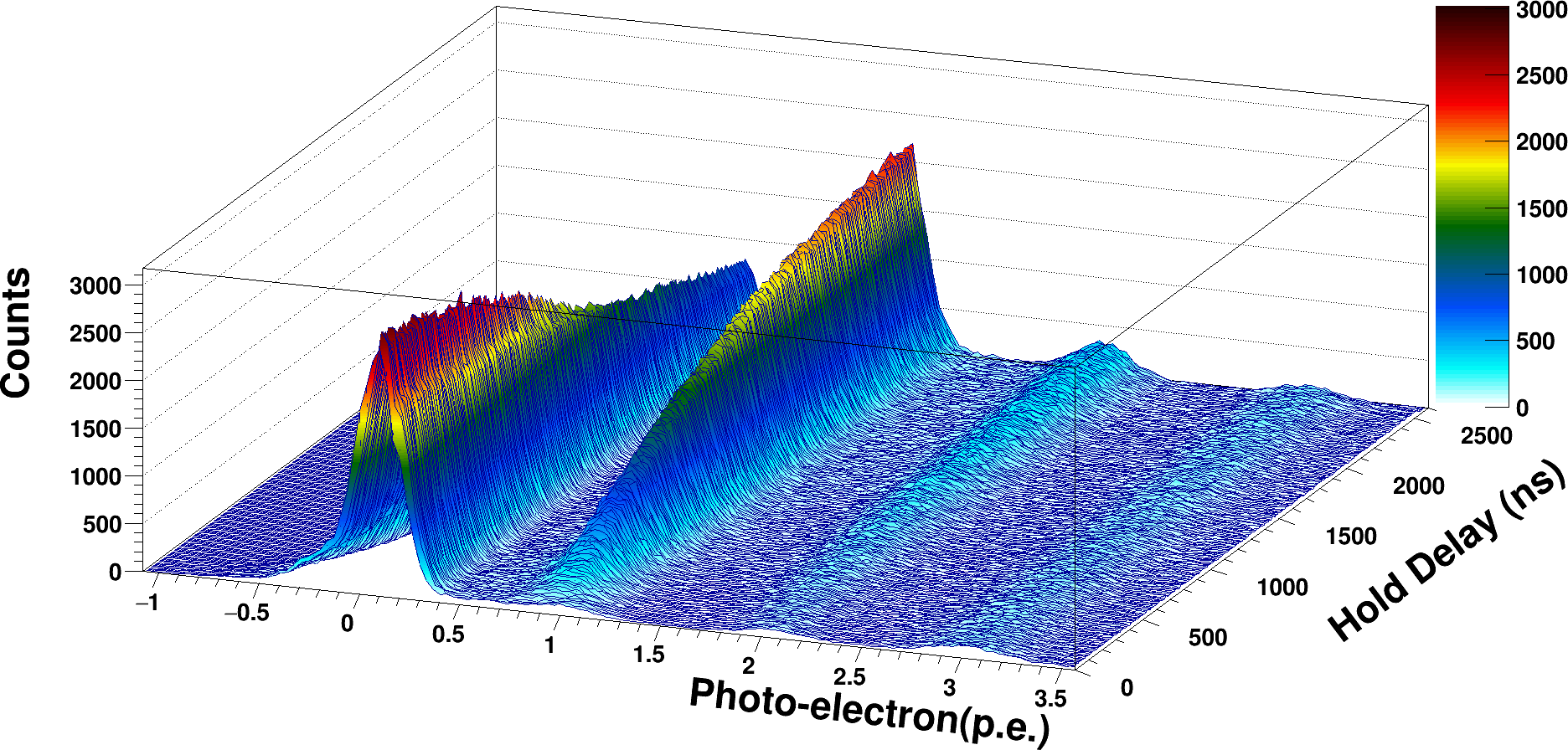}
\vspace{0.cm}
\caption{Three-dimensional pulse height spectrum as a function of the hold delay 
for real SiPM pulses in dark condition. 
It is evident that, by increasing the hold delay, it is possible to the decrease the number of counts in 
the pedestal peak in favour of the counts at multiple pe values.}
\label{fig6}
\end{figure*}

In peak detector working mode, the output is always the highest signal value in 
the time interval between the trigger and the rising edge of hold signal. 
The increase of the time interval between the trigger and the 
hold signal strongly affects therefore the pulsed height distribution.
In Fig.~\ref{fig5} it is clearly visible how the number of events 
accumulated in the first and second peak increases significantly with 
the hold delay.
In Fig.~\ref{fig6} we show the evolution of the dark count spectrum by varying the hold 
signal delay from 0 to 2550 ns in steps of ten ns.

\clearpage
\newpage
\section{Evaluation of the pixel gains using the peak detector}
\label{Evaluation of the pixel gains using the peak detector}

The peak detector behaviour shown in the previous section can be exploited for a precise 
evaluation of the SiPM pixel gain without using any 
external light source. In fact, just increasing the delay time between the trigger and 
the hold signal, it is possible to obtain dark count spectra with good statistical 
significance of the peaks at 1pe, 2pe, etc. 
 To check the validity of this method, we accumulated the 
dark pulse height spectra relative to an hold delay of 2550 ns. 
The dark spectrum is analysed with a multiple peaks Gaussian fit.
 The ADC values of the peaks are then plotted as a function of the 
number of photo-electrons and fitted (with the exclusion of the baseline peak)
to a straight line. The slope of the resulting 
fit line gives the pixel gains, while its extrapolation at 0pe provides the true baseline.
\\
As a consistency check, we applied the same method to the 
spectrum obtained by illuminating the SiPM with a ns pulsed blue-light-emitting diode (LED) setting the hold delay to 90 ns (see Fig.~\ref{fig8}). 

\begin{figure*}[h!!]
\centering
\includegraphics[angle=0, width=11cm]{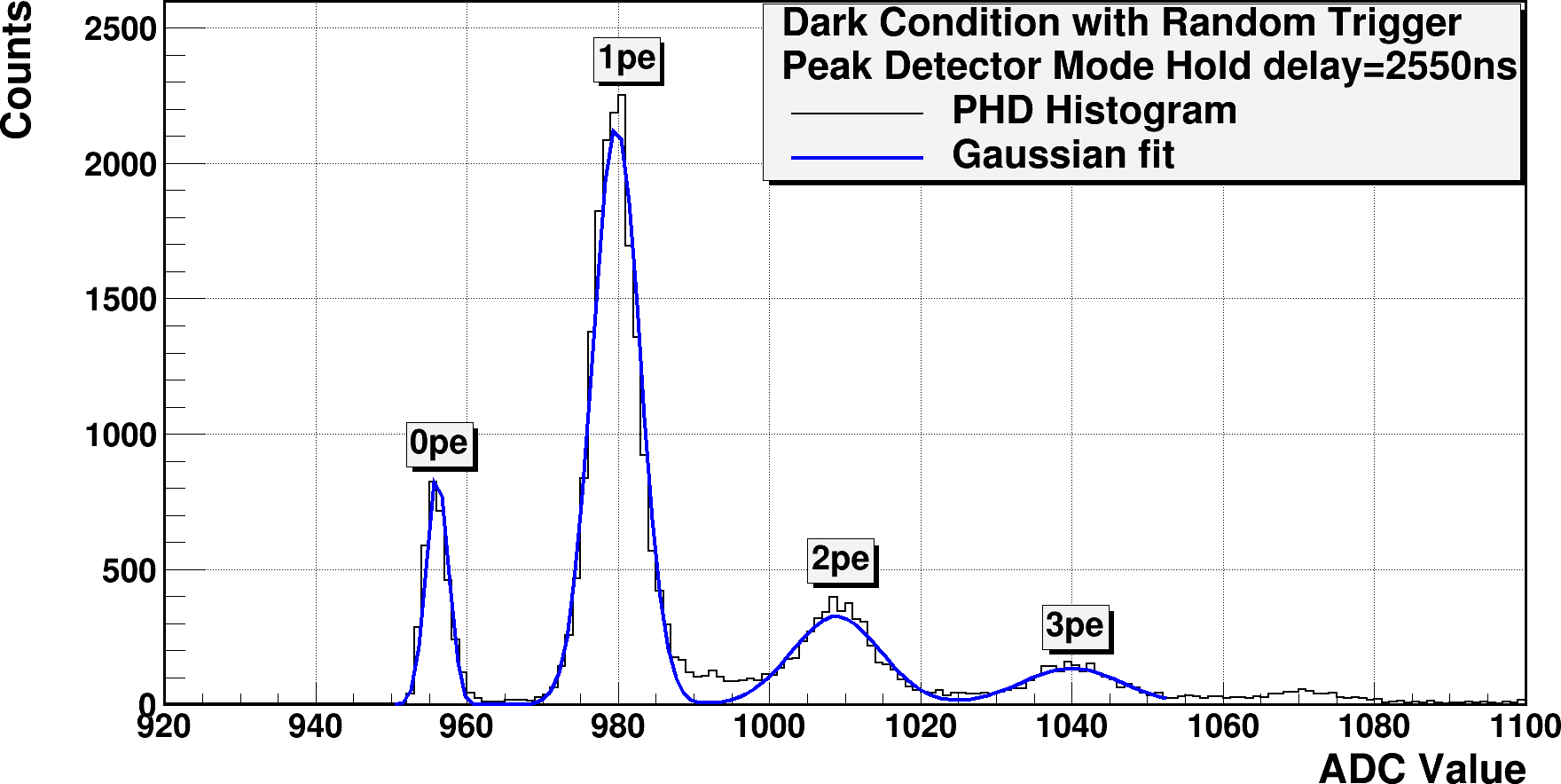}
\vspace{0.cm}
\caption{Single photo-electron peak spectrum taken  in peak detector mode with a Hold delay time of 2550 ns with SiPM pulses in dark condition.
The black curve represents the distribution of the peak detector output  and the blue curve shows 
the corresponding multiple peaks Gaussian fit.}
\label{fig7}
\end{figure*}

\begin{figure*}[h!!]
\centering
\includegraphics[angle=0, width=11cm]{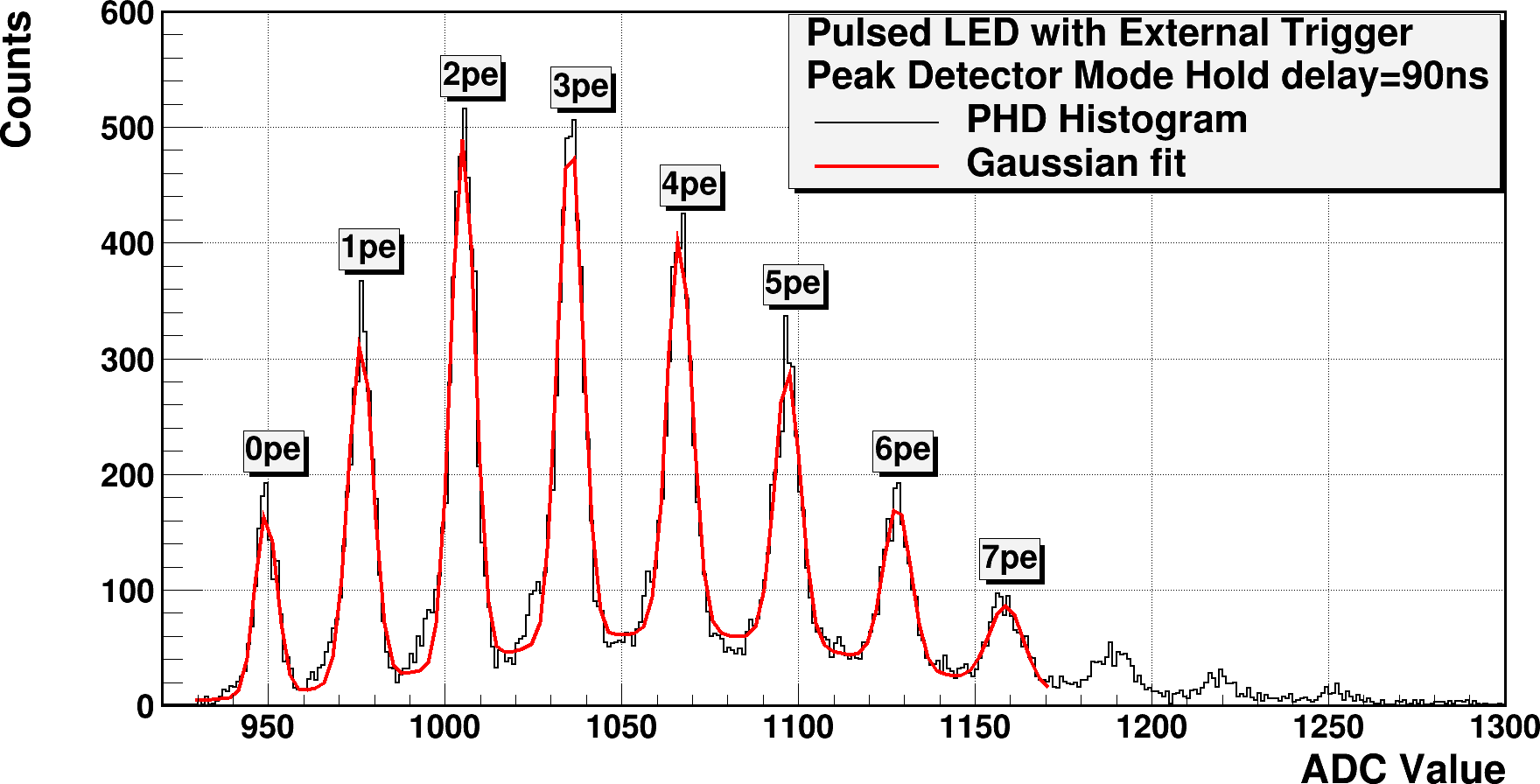}
\vspace{0.cm}
\caption{Pulse height spectrum acquired by illuminating the SiPM with 
a ns pulsed light source. The black curve represents the distribution of the peak detector output and the red curve shows 
the corresponding multiple peaks Gaussian fit.}
\label{fig8}
\end{figure*}
\begin{figure*}[h!!]
\centering
\includegraphics[angle=0, width=12cm]{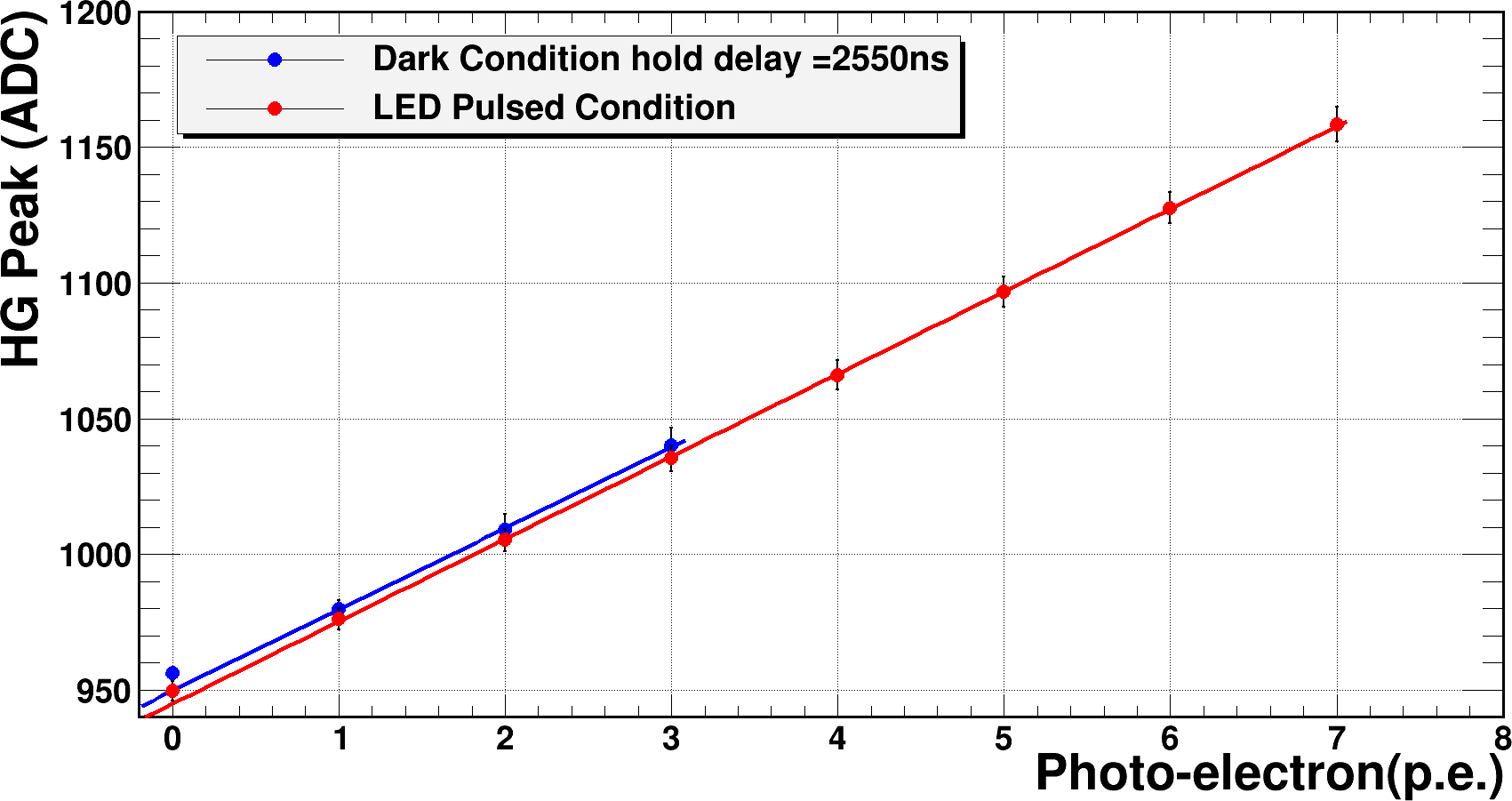}
\vspace{0.cm}
\caption{Values in ADC unit of the peaks of the spectrum distribution as a function of the number of photo-electrons, for the case of the spectrum 
obtained keeping the sensor in dark (blue points) and with the external pulsed light (red points).} 

\label{fig9}
\end{figure*}

 
\newpage
As shown in Fig.~\ref{fig9}, the pixel gain obtained from the dark counts (29.9$\pm$3.4 ADC/pe) 
is in good agreement, within the statistical errors, with the value achieved from the external pulsed light (30.4$\pm$0.9 ADC/pe).

Note that the value corresponding to the peak at 0pe, is not considered in the fit since it is, as expected, systematically higher than the baseline due to the positive bias 
induced by the peak detector working mode when no SiPM pulses are detected.
The true baseline can be however easily obtained as the value assumed by the fitting line at 0 pe.

The two methods present different advantages$\setminus$disadvantages that must be taken
 into account in designing a 
calibration system for a given experiment. 
The method with the external pulsed light provide rather smaller statistical errors than
the ''dark method'' because more peaks can be produced in the spectra distribution. 
On the other side, if  the scientific instrument has a large dimension, it 
may be difficult to deliver the ns light pulses to illuminate the detection plane without 
a significant temporal spreading of the photon pulses along their travel.
If excessive, the optical pulses broadening can therefore introduce systematic 
effects in the pulse height distributions and  systematic errors in the pixel gains measured with an external 
light source. 
On the contrary, the dark pulses generated by the cross-talk effect have the same identical temporal structure 
independently on their amplitude and, therefore, the method allows us to measure (although with a reduced number of peaks) the pixel gains
with a lower level of systematic .
An artificial pulsed light source able to provide large illumination levels is clearly necesssary to 
extend the gain measurement, from a few pe, up to the full working range of the instrument and to verify the responce 
linearity.    

.

\section{Conclusion}
We have shown that, by simply increasing the activaction time of peak detector working 
mode it is possible to obtain dark count spectra that allow us to calibrate the SiPM gain without an external 
light source. 
This calibration method is particularly useful for all experiments where the 
use of an external calibration light to illuminate the SiPM is not simple due to, e.g. the extended 
dimensions of the SiPM covered area.
In the context of Astri-Horn the proposed method will be used to complement and cross-check the 
results obtained with a pulsed light source illuminating the detection plane \cite{catalano18}.

\acknowledgments

This work is supported by the Italian Ministry of Education, University, 
and Research (MIUR) with funds specifically assigned to the Italian 
National Institute of Astrophysics (INAF) for the Cherenkov Telescope 
Array (CTA), and by the Italian Ministry of Economic Development (MISE) 
within the ''Astronomia Industriale'' program. We acknowledge support from 
the Brazilian Funding Agency FAPESP (Grant 2013/10559-5) and from the 
South African Department of Science and Technology through Funding 
Agreement 0227/2014 for the South African Gamma-Ray Astronomy Programme.\\ 
D.I., A.S., O.C., S.G. and T.M. thank S. Lombardi for his contribution as internal referee 
of the paper.

\end{document}